\documentclass{article}
\usepackage{tikz}
\usetikzlibrary{calc}
\usetikzlibrary{arrows}
 \usepackage{relsize}
 \usepackage[americanvoltage]{circuitikz}

\usepackage{amsmath}
\usepackage{amssymb}
\usetikzlibrary{calc}
\usepackage[americanvoltage]{circuitikz}
\usepackage{paralist}

\newcommand{\ppulp}[1]{\mathsf{#1}}

\newcommand{\ppulpm}[1]{\pmb{\ppulp{#1}}}

\newcommand*{\xdash}[1][3em]{\rule[0.5ex]{#1}{0.55pt}}

\newcommand{\vect}[1]{\mathbf{#1}}

\newcommand{\matr}[1]{\mathbf{#1}}
\newcommand{\abs}[1]{\left \lvert #1 \right\rvert }

\renewcommand{\Re}[1]{\mathbb{R}\mathrm{e}\left \{#1\right\} }
\renewcommand{\Im}[1]{\mathbb{I}\mathrm{m}\left \{#1\right\} }

\newcommand{\pref}[1]{(\ref{#1})}
\newcommand{\junk}[1] {}

\def\XXint#1#2#3{{\setbox0=\hbox{$#1{#2#3}{\int}$}
\vcenter{\hbox{$#2#3$}}\kern-.5\wd0}}

\newcommand*\widebar[1]{%
  \hbox{%
    \vbox{%
      \hrule height 0.5pt 
      \kern0.3ex
      \hbox{%
        \kern-0.05em
        \ensuremath{#1}%
        \kern-0.05em
      }%
    }%
  }%
}

\usepackage[margin=1in]{geometry}

\begin{document}

\title{Accurate Impedance Calculation for Underground and Submarine Power Cables using MoM-SO and a Multilayer Ground Model}

%
%
%
\author{Utkarsh~R.~Patel,
        and~Piero~Triverio \\
        Submitted to \emph{IEEE Transactions on Power Delivery} on March 16, 2015
\thanks{This work was supported in part by the KPN project "Electromagnetic transients in future power systems" (ref. 207160/E20) financed by the Norwegian Research Council (RENERGI programme) and by a consortium of industry partners led by SINTEF Energy Research: DONG Energy, EdF, EirGrid, Hafslund Nett, National Grid, Nexans Norway, RTE, Siemens Wind Power, Statnett, Statkraft, and Vestas Wind Systems.}
\thanks{U.~R.~Patel and P.~Triverio are with the Edward S. Rogers Sr. Department of Electrical and Computer Engineering, University of Toronto, Toronto, M5S 3G4 Canada (email: utkarsh.patel@mail.utoronto.ca, piero.triverio@utoronto.ca).}
}

\markboth{IEEE Transactions on Power Delivery}{IEEE Transactions on Power Delivery}%

\maketitle

\begin{abstract}
An accurate knowledge of the per-unit length impedance of power cables is necessary to correctly predict electromagnetic transients in power systems. In particular, skin, proximity, and ground return effects must be properly estimated.
In many applications, the medium that surrounds the cable is not uniform and can consist of multiple layers of different conductivity, such as dry and wet soil, water, or air. We introduce a multilayer ground model for the recently-proposed MoM-SO method, suitable to accurately predict ground return effects in such scenarios.  The proposed technique precisely accounts for skin, proximity, ground and tunnel effects, and is applicable to a variety of cable configurations, including underground and submarine cables. Numerical results show that the proposed method is more accurate than analytic formulas typically employed for transient analyses, and delivers an accuracy comparable to the finite element method (FEM). With respect to FEM, however, MoM-SO is over 1000 times faster, and can calculate the impedance of a submarine cable inside a three-layer medium in 0.10~s per frequency point.
\end{abstract}

\section{Introduction}
Electromagnetic transients, commonly induced by phenomena such as lightning and breakers operation, are a major source of power failures in today's power systems \cite{Ame13}.
In order to understand and mitigate these transients, power  engineers rely upon electromagnetic transient (EMT) simulation tools.
EMT tools need broadband models for all network components including cables, which are increasingly used by the power industry.
Broadband cable models \cite{Mar82, Nod96, Mor99} can be obtained only if the per-unit-length (p.u.l.) impedance and admittance of the cable are accurately known. 
In particular, such parameters must precisely account for the frequency-dependent behaviour of the cable caused by skin, proximity and ground effects.

Most EMT tools use analytic formulas \cite{Ame80, Wed73} to compute cable impedance.
Unfortunately, such formulas neglect proximity effects, which are strong in cables because of the tight spacing between conductors.
Numerical tools based on the finite-element method (FEM) \cite{Wei82, Cri89, B09, Hab13} or conductor partitioning \cite{Ame92, Com73, Dea87, Pag12} capture proximity effects inside the cable. 
However, these techniques require a fine discretization of the cable cross section in order to accurately capture skin effect, which leads to a large number of unknowns and long computational time.

In various fault scenarios, current flows through the ground that surrounds the cable.
Often, the ground surrounding the cable is non-uniform, and can only be modelled using multiple layers.
For example, submarine cables are surrounded by layers of air, sea, and seabed, all with different material properties.
In such scenario, the current flows inside each layer depending on each layer's conductivity.
Therefore, it is important to include effects of multilayer ground in cable models.

In most EMT tools, ground effects in underground cables are included through an approximation~\cite{Saad96} of Pollaczek's formulas~\cite{Pol26}.
These formulas can only model a surrounding medium made up of two half-spaces, with the top layer being air, and the bottom layer being either soil or water.
Pollaczek's formulas cannot accurately capture effects of a multilayer ground.
Several authors have developed analytic multilayer ground models \cite{Tsiamitros2008_1,Tsiamitros2008,pap2010}.
These formulas must be used in conjunction with either analytic formulas or FEM to account for cables' internal impedance \cite{Ame80}.
FEM-based techniques~\cite{yin89} can model multilayer ground, while including skin and proximity effects. 
However, their computational cost is very high because a very large region around the cable must be meshed in order to accurately estimate ground return impedance.

In \cite{TPWRD1, TPWRD2}, we presented an efficient technique called MoM-SO that computes proximity-aware p.u.l. impedance parameters of cables made up of round solid and hollow (tubular) conductors.
Our technique was later refined in~\cite{TPWRD3} to include the effects of single-layer ground and tunnels.
The efficiency of MoM-SO stems from the use of a surface admittance representation of the cable.
The surface admittance operator of the cable allows one to model the entire cable with a single equivalent current distribution on the boundary of the cable. 
The advantage of this is that we do not need to discretize the cross-section of the cable.
Instead, only the edges of the cable are discretized, which leads to significant computational savings.
With MoM-SO, ground effects are accounted for through the Green's function of the surrounding medium.
In this paper, we extend the technique to include a flexible multilayer ground model where the user can  specify an arbitrary number of soil layers with different conductivity. Numerical results demonstrate the superior accuracy of the proposed model, and how it can simplify the task of modeling a power cable in a non-uniform soil.

This paper is organized as follows. Section~\ref{sec:ProblemDefinition} defines the problem and the notation. 
In Sec.~\ref{sec:SurfaceOperatorCable}, we review the surface admittance formulation presented in~\cite{TPWRD3}. 
Then, in Sec.~\ref{sec:Multilayer}, multilayer ground effects are included via a multilayer Green's function.
The final expressions for p.u.l. impedance inclusive of skin, proximity, and ground effects are provided in Sec.~\ref{eq:pulparameters}.
Section~\ref{sec:Examples} validates the proposed technique against FEM, and shows the importance of an accurate modeling of proximity and ground effects.
\section{Problem Definition}
\label{sec:ProblemDefinition}

\begin{figure}
\centering
\begin{tikzpicture}[scale=0.85, every node/.style={scale=0.85}]
\draw [fill=white] (-1,3) rectangle (7,3.5) {};
\draw  [fill=black!10](-1,3) rectangle (7,0);
\draw [fill = black!5] (-1, 3.5) rectangle (7,4.5);
\draw [fill = white] (-1, -0.5) rectangle (7,0);
\draw [fill = black!5] (-1, -1.5) rectangle (7,-0.5);

\node at (-1,4) [right] {layer 1 ($\varepsilon_1, \mu_1$, $\sigma_1$)};

\draw [fill = black] (3,3.1) ellipse (0.04 and 0.04);
\draw  [fill = black] (3,3.25) ellipse (0.04 and 0.04);
\draw  [fill = black] (3,3.4) ellipse (0.04 and 0.04);

\draw [fill = black] (3,-0.1) ellipse (0.04 and 0.04);
\draw  [fill = black] (3,-0.25) ellipse (0.04 and 0.04);
\draw  [fill = black] (3,-0.4) ellipse (0.04 and 0.04);

\node at (-1,-1) [right] {layer L ($\varepsilon_L, \mu_L$, $\sigma_L$)};

\node at (5.8,-1.6) [above] {$y = y_L = -\infty$}; 
\node at (6, -0.5) [below] {$ y = y_{L-1}$};
\node at (6,-0.1) [above] {$y = y_{s}$};
\node at (6, 3) [below] {$y = y_{s-1}$};

\node at (6,3.4) [above] {$y = y_1$};
\node at (6,4.5) [below] {$y = y_0 = \infty$};

\node at (3,3.5) {};
\node at (3,2.8) {};

\draw  [ fill=white, densely dashed] (4,1.3) ellipse (1.28 and 1.28) node [left] {};
\draw  [fill=black!40, densely dashed] (3.5,1.85) ellipse (0.425 and 0.425) node {};
\node at (3.5,1.7) [above] {};

\draw  [fill = black!40,densely dashed] (4,0.6) ellipse (0.5 and 0.5) node [below] {} ;
\draw[fill=black](4,0.6) ellipse(0.01 and 0.01) node [above]{};
\node (v4) at (4,0.52)  [above] {};

\node at (4.5,2.2) [below] {};

\begin{scope}[shift = {(2.5,0)}]
\node (v2) at (5,2.8) {};
\node (v6) at (5,3.0) {};
\node (v5) at (4.8,3.0) {};
\node (v3) at (5,3.8) {\scriptsize ${\bf y}$};
\draw [-triangle 60] (v2) edge (v3);
\node (v4) at (5.8,3)  {};
\node(x1) at (5.7,3.1) {\scriptsize ${\bf x}$};
\draw [-triangle 60] (v5) edge (v4);
\end{scope}

\draw  [ fill=black] (v6) ellipse (0.06 and 0.06);
\draw  [ fill=black] (4,1.3) ellipse (0.02 and 0.02) node [left] (v8) {};
\node at (4.1,1.3) [left] {\scriptsize $(\hat{x}, \hat{y})$};

\node (v7) at (5.4,1.3) { };
\node at (4.7, 1.4) [below] {\scriptsize $\hat{a}$};
\draw[-stealth]   (4,1.3) -> (v7);

\node  at (5.2,0.6) {$\hat{c}$};
\node (center) at (4,1.3) {};
\node (center2) at (3.95,1.2) {};
\node (p1) at (5.4,1.3) {};
\node (p2) at (4.70,2.65) {};
\node(p12) at (4.47,2.2){};
\node (p3) at (4.2,1.65) {};
\node (p4) at (5.0, 2.3)  {};
\node (p5) at (5.4, 2.9) {};
\draw  [densely dashed] (center) -- (p2) {};
\draw  [->] (center2) -- (p12) {};

\node at (p3) [above] {\small $\hat{\rho}$};
\node at (p4) [right] {$\hat{\theta}$};
\draw [dashdotted] (center) -- (p1) {};
\draw [->] (p1) arc (0:60:1.4);

\node (p6) at (4, 0.6)  {};
\node (p66) at (3.9, 0.5)  {};
\node (p7) at (4.6, 0.6){};
\node(p8) at (4.5,1.08){};
\node(p9) at (4.25,1.4){};
\node(p10) at (4.75,1.5){};

\node at (4.1,0.85){\tiny$a_p$};

\draw [dashdotted] (p6) -- (p7){};
\draw [->] (p7) arc (0:45:0.6);
\draw [->] (p66) -- (p8){};
\draw  [fill=black] (p6) ellipse (0.03 and 0.03);

\node (p14) at (3.6,0.6) [left] {$c_p$};

\node (p15) at (4.47,0.8) [right] {\scriptsize $\theta_p$};
\node at (4,0.68) [below] {\tiny $(x_p, y_p)$};

\node at (-1,2.5) [right] {layer s ($\varepsilon_s, \mu_s, \sigma_s$)};
\node  (p17) at (2.05,1.5) [left] {conductors ($\varepsilon, \mu, \sigma$)};
\node (p18) at (3.2,1.8) {};
\node (p19) at (3.68,0.8) {};

\draw [->] (2.2,1.5)--(p18);
\draw [->] (2.2,1.5)--(p19);

\node (p20) at (-1,0.5) [right] {hole ($\hat{\varepsilon}, \hat{\mu}$)};
\node (p21) at (3.1,0.8) {};
\draw [->] (p20)--(p21);

\end{tikzpicture}
\caption{Cross-section of a simple cable made up of two solid conductors inside a hole. The medium surrounding the hole is modelled as $L$ horizontal layers with different conductivity, permittivity, and permeability.}
\label{fig:originalfig}
\end{figure}
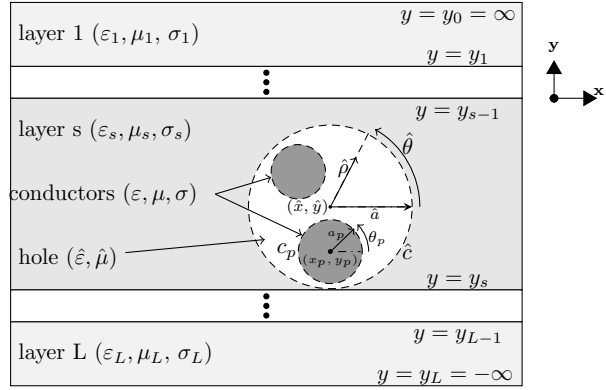

Our goal is to compute the p.u.l. impedance of underground and submarine power cables. 
In order to handle a large number of scenarios, we will devise a method applicable to cables with the following characteristics:
\begin{itemize}
\item any number of conductors, either solid or hollow, in arbitrary position and with arbitrary conductivity. Hollow conductors can be used to compactly represent an entire sheath or armour. Alternatively, individual strands can be also represented with solid conductors, owing to the excellent scalability of the proposed method;
\item a surrounding medium made by an arbitrary number of horizontal layers with arbitrary conductivity, as shown in Fig.~\ref{fig:originalfig}. This versatile ground model can be used to describe ground effects in a large number of scenarios. For example, in a shallow submarine cable, the electromagnetic field produced by certain transients propagates partially in air, partially in water, and partially in ground. 
Such scenario can be easily described in the adopted ground model using the top-most layer to represent air, a middle layer for water, and the bottom-most layer for seabed;
\item the presence of a hole or tunnel around the cable can be modelled;
\item multiple cable systems, possibly buried in different holes, can be also modelled in order to investigate mutual coupling.
\end{itemize}

For the sake of clarity, to describe the proposed method, we consider a cable made up of $P$ solid round conductors, all with the same conductivity $\sigma$, permittivity $\varepsilon$, and permeability $\mu$. 
Hollow conductors and conductors with different properties can also be handled, as shown in~\cite{TPWRD3}.
Conductor $p$ is centered at $(x_p, y_p)$, and has  outer radius  $a_p$. 
All $P$ conductors are placed inside a tunnel of radius $\hat{a}$ centered at $(\hat{x}, \hat{y})$, as illustrated in Fig~\ref{fig:originalfig}.
The hole consists of a lossless material with permittivity  $\hat{\varepsilon}$, and permeability $\hat{\mu}$.
The surrounding medium is made up of $L$ layers, as shown in Fig.~\ref{fig:originalfig}. Layer~$l$ has conductivity $\sigma_l$, permittivity $\varepsilon_l$, and permeability $\mu_l$. The top- and bottom-most layers, which are semi-infinite in the $y$-direction, are denoted as layer~1 and $L$, respectively. The layer in which the cable resides is denoted as layer~$s$. 

We are interested in computing the $P\times P$ p.u.l. resistance $\matr{R}(\omega)$ and inductance $\matr{L}(\omega)$ matrices that appear in the Telegrapher's equation
\begin{equation}
\frac{\partial \vect{V}}{\partial z} = - \left[\ppulpm{R}(\omega) + j \omega \ppulpm{L}(\omega) \right] \vect{I} \,,
\label{eq:Tel1}
\end{equation}
where vectors $\vect{V} = \begin{bmatrix} V_1 & V_2 & \hdots & V_P \end{bmatrix}^T$ and $\vect{I} = \begin{bmatrix} I_1 & I_2 & \hdots I_P \end{bmatrix}^T$ contain the potential $V_p$ and the current $I_p$ in each conductor, respectively.

\section{Surface Formulation for the Cable}
\label{sec:SurfaceOperatorCable}

In order to compute the cable impedance, we first apply the surface admittance formulation introduced in~\cite{DeZ05,TPWRD1}. This formulation reduces the complexity of the problem significantly, since a complex cable configuration can be described using a single equivalent current distribution. In this section, the surface formulation is briefly reviewed, and more details can be found in~\cite{TPWRD1,TPWRD2,TPWRD3}.

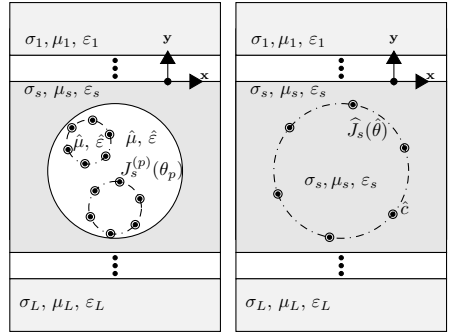
\begin{figure}
\centering
\begin{tikzpicture}[scale=0.7, every node/.style={scale=0.7}]
\draw [fill=white] (2,4) rectangle (6,0) node (v1) {};
\draw  [fill=black!10](2,3) rectangle (v1);

\draw [fill=white] (2,3) rectangle (6,3.5) {};
\draw  [fill=black!10](2,3) rectangle (6,-0.25);
\draw [fill = black!5] (2, 3.5) rectangle (6,4.5);
\draw [fill = white] (2, -0.75) rectangle (6,-0.25);
\draw [fill = black!5] (2, -1.75) rectangle (6,-0.75);

\draw [fill = black] (4,3.1) ellipse (0.04 and 0.04);
\draw  [fill = black] (4,3.25) ellipse (0.04 and 0.04);
\draw  [fill = black] (4,3.4) ellipse (0.04 and 0.04);

\draw [fill = black] (4,-0.35) ellipse (0.04 and 0.04);
\draw  [fill = black] (4,-0.5) ellipse (0.04 and 0.04);
\draw  [fill = black] (4,-0.65) ellipse (0.04 and 0.04);

\node at (3,3.75) {$\sigma_1, \mu_1$, $\varepsilon_1$};
\node at (3,2.8) {$\sigma_s$, $\mu_s$, $\varepsilon_s$};
\node at (3,-1.25) {$\sigma_L$, $\mu_L$, $\varepsilon_L$};

\draw  [ fill=white] (4,1.3) ellipse (1.28 and 1.28);
\draw  [dashdotted] [fill=white] (3.5,1.85) ellipse (0.42 and 0.42) node {$\hat{\mu}$, $\hat{\varepsilon}$};
\foreach \a in {20, 80,...,320} {
      \draw ({3.5+0.42*cos(\a)},{1.85+0.42*sin(\a)}) circle [radius=.08];
      \draw [fill=black] ({3.5+0.42*cos(\a)},{1.85+0.42*sin(\a)}) circle [radius=.04];
}


\draw  [dashdotted] [fill = white] (4,0.6) ellipse (0.5 and 0.5) node  {} ;
\foreach \a in {20, 80,...,320} {
      \draw ({4+0.5*cos(\a)},{0.6+0.5*sin(\a)}) circle [radius=.08];
      \draw [fill=black] ({4+0.5*cos(\a)},{0.6+0.5*sin(\a)}) circle [radius=.04];
}

\node at (4.5,2.2) [below] {$\hat{\mu}$, $\hat{\varepsilon}$};

\node at (4.5,2.2) [below] {};
\node at (4.7,1.35) { $J_s^{(p)}(\theta_p)$};


\foreach \a in {20, 80,...,320} {
}

\node (v2) at (5,2.8) {};
\node (v6) at (5,3.0) {};
\node (v5) at (4.8,3.0) {};
\node (v3) at (5,3.8) {\scriptsize $ {\bf y}$};
\draw [-triangle 60] (v2) edge (v3);

\node (v4) at (5.8,3)  {};
\node(x1) at (5.7,3.1) {\scriptsize ${\bf x}$};
\draw [-triangle 60] (v5) edge (v4);

\draw  [ fill=black] (v6) ellipse (0.06 and 0.06);


\end{tikzpicture}
\begin{tikzpicture}[scale=0.7, every node/.style={scale=0.7}]
\draw [fill=white] (2,4) rectangle (6,0) node (v1) {};
\draw  [fill=black!10](2,3) rectangle (v1);

\draw [fill=white] (2,3) rectangle (6,3.5) {};
\draw  [fill=black!10](2,3) rectangle (6,-0.25);
\draw [fill = black!5] (2, 3.5) rectangle (6,4.5);
\draw [fill = white] (2, -0.75) rectangle (6,-0.25);
\draw [fill = black!5] (2, -1.75) rectangle (6,-0.75);

\draw [fill = black] (4,3.1) ellipse (0.04 and 0.04);
\draw  [fill = black] (4,3.25) ellipse (0.04 and 0.04);
\draw  [fill = black] (4,3.4) ellipse (0.04 and 0.04);

\draw [fill = black] (4,-0.35) ellipse (0.04 and 0.04);
\draw  [fill = black] (4,-0.5) ellipse (0.04 and 0.04);
\draw  [fill = black] (4,-0.65) ellipse (0.04 and 0.04);

\node at (3,3.75) {$\sigma_1, \mu_1$, $\varepsilon_1$};
\node at (3,2.8) {$\sigma_s$, $\mu_s$, $\varepsilon_s$};
\node at (3,-1.25) {$\sigma_L$, $\mu_L$, $\varepsilon_L$};


\draw  [dashdotted, fill=black!10] (4,1.3) ellipse (1.28 and 1.28)  node [below] {$\sigma_s, \mu_s$, $\varepsilon_s$};
\foreach \a in {20, 80,...,320} {
      \draw ({4+1.28*cos(\a)},{1.3+1.28*sin(\a)}) circle [radius=.08];
      \draw [fill=black] ({4+1.28*cos(\a)},{1.3+1.28*sin(\a)}) circle [radius=.04];
}

\node at (4.5, 2.1) {$\widehat{J}_s(\hat{\theta})$};



\node (v2) at (5,2.8) {};
\node (v6) at (5,3.0) {};
\node (v5) at (4.8,3.0) {};
\node (v3) at (5,3.8) {\scriptsize ${\bf y}$};
\draw [-triangle 60] (v2) edge (v3);

\node (v4) at (5.8,3)  {};
\node (x1) at (5.7,3.1) {\scriptsize $ {\bf x}$};
\draw [-triangle 60] (v5) edge (v4);

\draw  [ fill=black] (v6) ellipse (0.06 and 0.06);

\node  at (5.2,0.6) {$\hat{c}$};

\end{tikzpicture}
\caption{Left panel: simplified problem after the application of the equivalence theorem to the conductors. Right panel: problem after the application of the equivalence theorem to the hole.
}
\label{fig:replaceconfig}
\end{figure}

\subsection{Surface Admittance for Round Conductors}

We first expand the electric field on the boundary of each conductor using a truncated Fourier series. For the $p$-th conductor, this expansion is
\begin{equation}
E_z(\theta_p) = \sum_{n=-N_p}^{N_p} E_n^{(p)} \operatorname{e}^{jn\theta_p} \,,
\label{eq:Eexp1}
\end{equation}
where $E_n^{(p)}$ are the Fourier coefficients of the electric field. The azimuthal coordinate $\theta_p$ is used to trace the boundary of conductor $p$, as shown in Fig.~\ref{fig:originalfig}.
In \pref{eq:Eexp1}, $N_p$ denotes the number of harmonics used to represent the electric field distribution. For most practical cases, $N_p = 4$ is sufficient~\cite{TPWRD1, ipst2013}. 
The coefficients of the electric field on all the conductors are collected into a global vector $\vect{E} = \begin{bmatrix} E_{-N_1}^{(1)} & \hdots &  E_{N_1}^{(1)} & \hdots & E_{-N_P}^{(P)} & \hdots &  E_{N_P}^{(P)}\end{bmatrix}^T$.
Next, we replace all conductors with the surrounding hole medium, and introduce an equivalent current density on their boundary in order to keep the electric field outside the conductors unchanged. This transformation, illustrated in the left panel of Fig.~\ref{fig:replaceconfig}, is enabled by the equivalence theorem~\cite{Bal05}.
As for the electric field in~\pref{eq:Eexp1}, the equivalent current for each conductor is  expanded in a truncated Fourier series
\begin{equation}
J_s^{(p)}(\theta_p) = \frac{1}{2\pi a_p} \sum_{n=-N_p}^{N_p} J_n^{(p)} \operatorname{e}^{jn\theta_p} \,.
\label{eq:Jexpan1}
\end{equation}
The current coefficients $J_n^{(p)}$ for all conductors are collected into a column vector

\noindent $\vect{J} = \begin{bmatrix} J_{-N_1}^{(1)} & \hdots & J_{N_1}^{(1)} & \hdots & J_{-N_P}^{(P)} & \hdots & J_{N_P}^{(P)} \end{bmatrix}^T$.
As shown in~\cite{DeZ05,TPWRD1}, the equivalent current coefficients $\vect{J}$ are related to $\vect{E}$ by a surface admittance operator $\matr{Y}_s$
\begin{equation}
\vect{J} = \matr{Y}_s \vect{E} \,.
\label{eq:surf1}
\end{equation}
This compact relation, which can be derived analytically for both solid~\cite{DeZ05, TPWRD1} and hollow~\cite{TPWRD2} round conductors, is sufficient to completely describe the conductors' influence on the cable impedance.

\subsection{Surface Admittance for a Cable-Hole System}

The geometry of the problem can be further simplified by applying the equivalence theorem a second time to the hole boundary. We first introduce the Fourier expansion of the magnetic vector potential on the hole boundary
\begin{equation}
\widehat{A}_z(\hat{\theta}) = \sum_{n=-\widehat{N}}^{\widehat{N}} \widehat{A}_n \operatorname{e}^{jn\hat{\theta}} \,.
\label{eq:BC1}
\end{equation}
Using the equivalence theorem~\cite{Bal05}, we replace the entire hole with the surrounding medium, i.e. having the same material parameters of layer $s$. An equivalent current density $\widehat{J}_s(\hat{\theta})$ is introduced on the hole boundary, and expressed in Fourier series
\begin{equation}
\widehat{J}_s(\hat{\theta}) = \frac{1}{2\pi \hat{a}} \sum_{n=-\widehat{N}}^{\widehat{N}} \widehat{J}_n \operatorname{e}^{jn\hat{\theta}} \,,
\label{eq:JHat1}
\end{equation}
where $\widehat{N}$ controls the number of harmonics used to represent the cable-hole system. 
The hole equivalent current $\widehat{J}_s(\hat{\theta})$ is found with the following relation~\cite{TPWRD3}
\begin{equation}
\widehat{\vect{J}} = \widehat{\matr{Y}}_s \widehat{\vect{A}} + \vect{T} \vect{J} \,,
\label{eq:surf2}
\end{equation}
where $\widehat{\vect{J}} = \begin{bmatrix} \widehat{J}_{-\widehat{N}} & \hdots & \widehat{J}_{\widehat{N}}\end{bmatrix}^T$ and $\widehat{\vect{A}} = \begin{bmatrix} \widehat{A}_{-\widehat{N}} & \hdots & \widehat{A}_{\widehat{N}} \end{bmatrix}^T$. From~\pref{eq:surf2}, we can recognize two contributions to the equivalent hole current.
The first term in~\pref{eq:surf2} is analogous to the surface admittance operator of round conductors~\pref{eq:surf1}, and models an empty hole. The second term in~\pref{eq:surf2} accounts for the presence of the cable conductors inside the hole. The transformation matrix $\matr{T}$, maps the equivalent conductor currents~\pref{eq:Jexpan1} onto the boundary of the hole. Analytic expressions for $\widehat{\matr{Y}}_s$ and $\matr{T}$ can be found in~\cite{TPWRD3}.

\section{Multilayer Ground Model}
\label{sec:Multilayer}

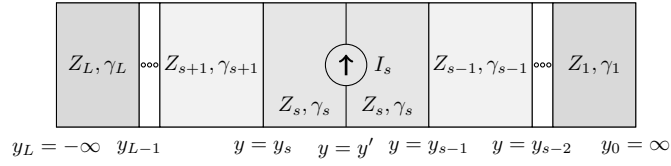
\begin{figure}[t]
\centering
\begin{tikzpicture}[scale=0.55, every node/.style={scale=0.8}]
\draw  (-7,3) rectangle (7,0);

\node (p1) at (-7.0,3) {};
\node (p2) at (-5,0) {};
\node (p3) at (-4.5,3) {};
\node (p4) at (-2,0) {};
\node (p144) at (2, 3) {};
\node (p133) at (4.5, 0) {};
\node (p122) at (5, 3) {};
\node (p111) at (7, 0) {};

\draw [fill = black!15] (p1) rectangle (p2);
\draw [fill = black!0] (p2) rectangle (p3);
\draw [fill = black!5] (p3) rectangle (p4);
\draw [fill = black!10] (p144) rectangle (p4);
\draw [fill = black!5] (p144) rectangle (p133);
\draw [fill = black!0] (p133) rectangle (p122);
\draw [fill = black!15] (p111) rectangle (p122);

\node at (0.5,1.5) [right]{$I_{s}$};
\draw  (0,1.5) ellipse (0.5 and 0.5);
\draw  (-4.75,1.5) ellipse (0.05 and 0.05);
\draw  (-4.6,1.5) ellipse (0.05 and 0.05);
\draw  (-4.9,1.5) ellipse (0.05 and 0.05);
\draw  (4.75,1.5) ellipse (0.05 and 0.05);
\draw  (4.6,1.5) ellipse (0.05 and 0.05);
\draw  (4.9,1.5) ellipse (0.05 and 0.05);

\draw  (0,2) edge (0,3);
\draw  (0,1) edge (0,0);
\draw [line width=1, ->] (0,1.25) -- (0,1.75){};

\node at (2,-0.5) {$y = y_{s-1}$};
\node at (0,-0.5) {$y = y'$};
\node at (-2,-0.5) {$y = y_{s}$};
\node at (4.5,-0.5){$y = y_{s-2}$};
\node at (7,-0.5)  {$y_0 = \infty$};
\node at (-5,-0.5) {$y_{L-1}$};
\node at (-7,-0.5) {$y_L =-\infty$};

\node at (-6,1.5) {$Z_L, \gamma_L$};
\node at (-3.25,1.5) {$Z_{s+1}, \gamma_{s+1}$};
\node at (-1,0.5) {$Z_s, \gamma_s$};
\node at (1,0.5) {$Z_s, \gamma_s$};
\node at (3.25,1.5) {$Z_{s-1}, \gamma_{s-1}$};
\node at (6.0,1.5) {$Z_{1}, \gamma_{1}$};

\end{tikzpicture}
\vspace{-5mm}
\caption{Equivalent transmission line model used to represent the multilayer surrounding medium. Each layer is modelled as a segment of transmission line.}
\label{fig:TMmodel}
\end{figure}

Using the equivalence theorem, we have restored the homogeneity of the problem in each layer, as shown in the right panel of Fig.~\ref{fig:replaceconfig}. In this equivalent configuration, the Green's function can be conveniently used to relate currents and electromagnetic fields, and determine the cable impedance. We relate the vector potential and equivalent current on the hole boundary through the magnetic vector potential integral equation~\cite{Bal05}
\begin{equation}
\widehat{{A}}_z(\hat{\theta}) = - \mu_s \int_0^{2\pi} \widehat{J}_s(\hat{\theta} ') {G}_{g}\left(\widehat{\vect{r}}(\hat{a}, \hat{\theta}),  \widehat{\vect{r}}(\hat{a}, \hat{\theta}')\right) \hat{a} d \hat{\theta}' \,,
\label{eq:outerefie}
\end{equation}
where the integral kernel $G_g(.,.)$ is the Green's function of the multilayer medium shown in the right panel of Fig.~\ref{fig:replaceconfig}. Position vector $\widehat{\vect{r}}(\hat{a}, \hat{\theta})$ traces the contour of the hole.

\subsection{Multilayer Green's function}

The Green's function $G_g$ is related to the $z$-oriented magnetic vector potential ${\cal A}_z(x,y)$ due to a $z$-oriented point source placed at $(x',y')$. 
The Green's function of a multilayer medium can be found from the nonhomogeneous Helmholtz equation~\cite{Fache93, Mic97}
\begin{equation}
\nabla^2{G}_g(x,y) + k_s^2 {G}_g(x,y) = \delta(x-x',y-y')\,
\label{eq:HelmholtzEquation2}
\end{equation}
in layer~$s$, where the $\delta$-function is centered at $(x', y')$ and $k_s = \omega\mu_s \sqrt{\omega \varepsilon_s  - j \sigma_s}$ is the wave number inside the layer.
In all other layers, the Green's function satisfies the homogeneous Helmholtz equation 
\begin{equation}
\nabla^2{G}_g(x,y) + k_l^2 {G}_g(x,y) = 0\,,
\label{eq:HelmholtzEquation}
\end{equation}
where $k_l = \omega\mu_l \sqrt{\omega \varepsilon_l - j \sigma_l}$ is the wavenumber inside layer $l$.
To solve \pref{eq:HelmholtzEquation2} and~\pref{eq:HelmholtzEquation}, we apply the Fourier transform with respect to $x$  to obtain
\begin{equation}
\frac{\partial^2}{\partial y^2} \widetilde{G}_g(\beta_x, y) - (\beta_x^2 - k_s^2) \widetilde{G}_g(\beta_x, y) = e^{j\beta_x x'} \delta(y-y')\,
\label{eq:HelmholtzSpectral2}
\end{equation}
in layer $s$, and
\begin{equation}
\frac{\partial^2}{\partial y^2} \widetilde{G}_g(\beta_x, y) - (\beta_x^2 - k_l^2) \widetilde{G}_g(\beta_x, y) = 0\,
\label{eq:HelmholtzSpectral1}
\end{equation}
in layer $l \ne s$, where 
\begin{align}
\widetilde{G}_{g}(\beta_x, y) &= \int_{-\infty}^{\infty} {G}_g(x,y) e^{j\beta_x x} dx\,.
\end{align}

It can be shown \cite{Fache93, Mic97} that solving \pref{eq:HelmholtzSpectral2} and \pref{eq:HelmholtzSpectral1} is equivalent to solving the equivalent transmission line (TL) circuit shown in Fig.~\ref{fig:TMmodel}.
In this TL model, each layer of the background medium is modelled as a segment of TL, with length equal to the height of the layer, characteristic impedance $Z_l = \left( {\beta_x^2 - k_l^2}\right)^{-1/2} \,,$
and propagation constant $\gamma_l = \sqrt{\beta_x^2 - k_l^2}$.
The point source in \pref{eq:HelmholtzEquation2} is modelled as a current source $I_{s} = e^{j\beta_x x'}$ located at $y = y'$.
In the TL model of Fig.~\ref{fig:TMmodel}, the voltage along the line is equal to the spectral domain Green's function $\widetilde{G}_g(\beta_x, y)$. 

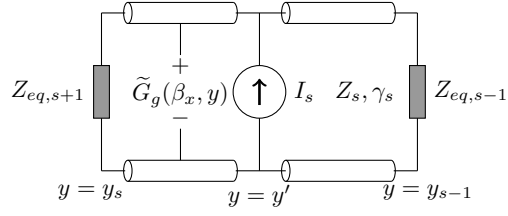
\begin{figure}
\centering
\begin{tikzpicture}[scale=0.7, every node/.style={scale=0.9}]

\draw [fill = black!40] (-3.15, 1.0) rectangle (-2.85, 2);
\node at (-3.15, 1.5) [left] {$Z_{eq, s+1}$};

\draw [fill = black!40] (3.15, 1.0) rectangle (2.85, 2);
\node at (3.15, 1.5) [right] {$Z_{eq, s-1}$};




\node (p13) at (0,0) {};
\node (p14) at (0,3){};
\node (p15) at (0,1) {};
\node (p16) at (0,2) {};
\node (p17) at (0, 1.5) {};
\node (p18) at (0,1.25){};
\node (p19) at (0,1.75) {};

\draw  (p17) ellipse (0.5 and 0.5);
\draw  (p16.center) edge (p14.center);
\draw  (p13.center) edge (p15.center);
\draw [line width=1, ->] (p18.center) -- (p19.center){};


\node at (-3.2,-0.1) [below] {$y = y_{s}$};
\node at (0,-0.1) [below] {$y = y'$};
\node at (3.2,-0.1) [below] {$y = y_{s-1}$};

\node at (2,1.5) {$Z_s, \gamma_s$};

\node at (0.5,1.5) [right]{$I_{s}$};

\draw (0.5,3) ellipse (0.08 and 0.2);
\draw (0.5, 3.2) -- (2.5,3.2);
\draw (0.5, 2.8) -- (2.5,2.8);
\draw (2.5,2.8) arc (-90:90:0.08 and 0.2);

\begin{scope}[shift = {(0,-3)}]
\draw (0.5,3) ellipse (0.08 and 0.2);
\draw (0.5, 3.2) -- (2.5,3.2);
\draw (0.5, 2.8) -- (2.5,2.8);
\draw (2.5,2.8) arc (-90:90:0.08 and 0.2);
\end{scope}

\begin{scope}[shift = {(-3,-3)}]
\draw (0.5,3) ellipse (0.08 and 0.2);
\draw (0.5, 3.2) -- (2.5,3.2);
\draw (0.5, 2.8) -- (2.5,2.8);
\draw (2.5,2.8) arc (-90:90:0.08 and 0.2);
\end{scope}

\begin{scope}[shift = {(-3,0)}]
\draw (0.5,3) ellipse (0.08 and 0.2);
\draw (0.5, 3.2) -- (2.5,3.2);
\draw (0.5, 2.8) -- (2.5,2.8);
\draw (2.5,2.8) arc (-90:90:0.08 and 0.2);
\end{scope}

\draw (-0.45,0) -- (0.5,0);
\draw (-0.45,3) -- (0.5,3);
\draw (2.55,0) -- (3.0,0);
\draw (2.55,3) -- (3.0,3);
\draw (-2.55,0) -- (-3.0,0);
\draw (-2.55,3) -- (-3.0,3);
\draw (3,0) -- (3.0,1);
\draw (3,2) -- (3.0,3);
\draw (-3,0) -- (-3.0,1);
\draw (-3,2) -- (-3.0,3);

\node at (-1.5, 1.5) {$\widetilde{G}_g(\beta_x, y)$};
\node at (-1.5, 2) {$+$};
\node at (-1.5,1) {$-$};
\draw (-1.5,2.2) -- (-1.5,2.8);
\draw (-1.5,0.2) -- (-1.5,.8);

\end{tikzpicture}
\caption{Simplified transmission line model assuming that both source coordinates $(x',y')$ and observation coordinates $(x,y)$ are in layer~s.}
\label{fig:simplifiedTMmodel}
\end{figure}

To evaluate \pref{eq:outerefie}, we only require voltage in layer~$s$ of the TL model in Fig.~\ref{fig:TMmodel}.
Therefore, we can simplify the model in Fig.~\ref{fig:TMmodel} by replacing layers $s+1, \hdots, L$ by an equivalent impedance $Z_{eq, s+1}$ and layers $1, \hdots, s-1$ by an equivalent impedance $Z_{eq, s-1}$, as shown in Fig.~\ref{fig:simplifiedTMmodel}. 
Both equivalent impedances are easily calculated using TL input impedance formulas found in most electromagnetic and TL theory textbooks~\cite{Che89}.
The solution of the equivalent circuit in Fig.~\ref{fig:simplifiedTMmodel} is \cite{Tes97}
\begin{align}
\widetilde{G}_g&(\beta_x, y) = \left( \frac{Z_s I_s}{2} \right)  e^{\left(-\abs{y - y'} \gamma_s \right) } \label{eq:ch3_AzSpectralVal1} \\
&+ \left(\frac{Z_s I_s}{2}  \right) \left [ \frac{1}{1 - \Gamma_R \Gamma_L e^{-2 (y_{s-1} - y_{s}) \gamma_s }} \right] \cdot \nonumber \\
&  \bigg [ \Gamma_L e^{(2y_{s} - y' - y ) \gamma_s }  +\Gamma_R \Gamma_L e^{ \left(2y_{s} - 2y_{s-1} + y' - y \right) \gamma_s }  \nonumber \\
   &+ \Gamma_R e^{ (-2y_{s-1} + y +y')\gamma_s } + \Gamma_R \Gamma_L e^{ \left(2 y_{s} - 2y_{s-1} + y - y' \right) \gamma_s  } \bigg] \nonumber \,,
\end{align} 
where $\Gamma_L$ and $\Gamma_R$ are the reflection coefficients
\begin{align}
\Gamma_L = \frac{Z_{eq,s+1} - Z_s}{Z_s + Z_{eq,s+1}}\,, \\
\Gamma_R = \frac{Z_{eq,s-1} - Z_s}{Z_s + Z_{eq,s-1}}\,.
\end{align}

Finally, we take the inverse Fourier transform of \pref{eq:ch3_AzSpectralVal1} to obtain the desired Green's function
\begin{equation}
{G}_{g}(x,y) = \frac{1}{2\pi} \int_{-\infty}^{\infty} \widetilde{G}_g(\beta_x, y) e^{-j\beta_x x} d\beta_x\,,
\label{eq:MultilayerGreensfunction}
\end{equation}
where the integral can be evaluated numerically.

\subsection{Discretized Integral Equation}

Next, we substitute the Fourier series expansion of vector potential \pref{eq:BC1} and equivalent hole current \pref{eq:JHat1}, as well as the multilayer Green's function \pref{eq:MultilayerGreensfunction} into \pref{eq:outerefie}. 
The resulting integral equation is discretized using the method of moments~\cite{Har61}, obtaining
\begin{equation}
\widehat{\vect{A}} = - \mu_s \matr{G}_g \widehat{\vect{J}} \,,
\label{eq:EFIEhalf}
\end{equation}
where $\matr{G}_g$ is a matrix representation of the discretized multilayer Green's function \cite{TPWRD1}.
By substituting \pref{eq:surf2} into \pref{eq:EFIEhalf}, we obtain the magnetic vector potential coefficients
\begin{equation}
\widehat{\matr{A}} = -\mu_s \left(\matr{1} + \mu_s \matr{G}_g \widehat{\matr{Y}}_s \right)^{-1} \matr{G}_g \matr{T} \matr{J}\,.
\end{equation}

\section{Computation of Cable Impedance}
\label{eq:pulparameters}

To compute the p.u.l. impedance parameters we need the electric field on the conductors, which can be found by evaluating the electric field integral equation \cite{Bal05}
\begin{equation}
E_z (\vect{r}_p(\theta_p)) = - j\omega \widehat{\cal A}_z - \frac{\partial V}{\partial z}\,,
\label{eq:efie1}
\end{equation}
on the contour of each conductor. In~\pref{eq:efie1}, the position vector $\vect{r}_p(\theta_p)$ traces the contour of the $p$-th conductor.
Using the same steps outlined in \cite{TPWRD3}, we finally arrive at the p.u.l. resistance and inductance of the cable
\begin{align}
\ppulpm{R}(\omega)&= \Re { \left( \matr{U}^T\left( \matr{1} - j \omega \matr{Y}_s \Psi  \right) ^{-1} \matr{Y}_s \vect{U} \right)^{-1}} \,, \label{eq:R} \\
\ppulpm{L}(\omega) &= \omega^{-1}\Im { \left( \matr{U}^T\left( \matr{1} - j \omega\matr{Y}_s \Psi  \right) ^{-1} \matr{Y}_s \vect{U} \right)^{-1}} \label{eq:L} \,,
\end{align}
where
\begin{equation}
\matr{\Psi} =  \widehat{\matr{H}} \matr{D}_1  \left[ {\mu}_s \left(\vect{1} + {\mu}_s \matr{G}_g \widehat{\matr{Y}}_s \right)^{-1} \matr{G}_g \matr{T}  - \hat{\mu} \widehat{\matr{G}}_0 \right] + \hat{\mu} \widehat{\matr{G}}_c  \,,
\label{eq:Psi}
\end{equation}
and matrices $\widehat{\matr{H}}$, $\matr{D}_1$, $\matr{1}$, $\matr{U}$, $\widehat{\matr{G}}_0$, and $\widehat{\matr{G}}_c$ are defined in \cite{TPWRD3}.


\section{Numerical Examples}
\label{sec:Examples}

\begin{figure}
\begin{center}
\begin{tikzpicture}[scale=0.30, every node/.style={scale=0.32}]

\tikzstyle{line} = [draw, -latex', <->];
\draw [fill=black!20] (0,0.5) rectangle (16,9.5);
\draw [fill=black!5] (0,6.8) rectangle (16,9.5);
\draw [fill = white] (0, 9.5) rectangle (16, 11);

\draw [fill=white]  (2.5,3.5) ellipse (2 and 2);
\draw [fill=black!80] (2.5,3.5) ellipse (1.8 and 1.8);
\draw [fill=white] (2.5,3.5) ellipse (1.7 and 1.7);
\draw [fill=black!50] (2.5,3.5) ellipse (1.3 and 1.3);

\draw [fill=white]  (7.0,3.5) ellipse (2 and 2);
\draw [fill=black!80] (7.0,3.5) ellipse (1.8 and 1.8);
\draw [fill=white] (7.0,3.5) ellipse (1.7 and 1.7);
\draw [fill=black!50] (7.0,3.5) ellipse (1.3 and 1.3);

\draw [fill=white]  (11.5,3.5) ellipse (2 and 2);
\draw [fill=black!80] (11.5,3.5) ellipse (1.8 and 1.8);
\draw [fill=white] (11.5,3.5) ellipse (1.7 and 1.7);
\draw [fill=black!50] (11.5,3.5) ellipse (1.3 and 1.3);

\node (p1) at (14,3.5) {};
\node (p2) at (14,6.8) {};
\path [line] (p1) -- (p2) {};

\node (p3) at (2.5,5.8) {};
\node (p4) at (7.0,5.8) {};
\path [line] (p3) -- (p4)  {};

\draw [<->] (14, 9.5) -- (14, 7);
\node at (14, 8.25) [right]{\huge 10~m};

\node (p5) at (4.75,5.8) [above]{\huge D};
\node (p6) at (14,5.0) [right] {\huge 1~m};

\node at (8,0.95) {\huge seabed ($\varepsilon_3 = 15 ~\varepsilon_0, \mu_3 = \mu_0, \sigma_3 = 0.05~{\rm S/m}$)};

\node at (8,10.25) {\huge air ($\varepsilon_0 , \mu_0$)};
\node at (8,8.25) {\huge sea ($\varepsilon_2= 81~ \varepsilon_0, \mu_2 = \mu_0, \sigma_2 = 5~{\rm S/m}$)};

\end{tikzpicture}
\caption{System of three single core cables used for validation in Sec.~\ref{sec:Examples}.}
\label{fig:Validation1}
\end{center}
\end{figure}
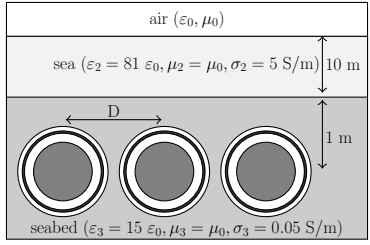

We now present two numerical examples to validate the proposed technique and to emphasize the need to include proximity-aware cable and multilayer ground models in EMT simulations.

\subsection{Example \#1 - Three Tightly-Spaced  Single-Core Cables}
\label{sec:example1}

\subsubsection{Cable Geometry and Material Parameters}

\begin{table}[t]
\centering
\caption{Single core cables of Sec.~\ref{sec:Examples}: geometrical and material parameters}
\begin{tabular}{|c|c|}
\hline
Core & Outer diameter = 39~mm, $\rho = 3.365\cdot 10^{-8} {\rm~\Omega \cdot m}$ \\ \hline
Insulation & $t = 18.25 {\rm~mm} $, $\varepsilon_r = 2.85$ \\ \hline
Sheath & $t = 0.22{\rm~mm}$, $\rho = 1.718\cdot 10^{-8} {\rm~\Omega \cdot m}$ \\ \hline
Jacket & $t = 4.53{\rm~mm}$, $\varepsilon_r = 2.51$ \label{tab:SCparameter} \\ \hline
\end{tabular}
\label{tab:parameters}
\end{table}

The first example consists of three single-core cables (SC) buried under a shallow sea, as shown in Fig.~\ref{fig:Validation1}.
The geometrical and material properties of the cables are listed in Table~\ref{tab:parameters}.
The surrounding medium is modelled using $L=3$ layers. The top layer models air. The second layer models the sea,  with height of 10~m, conductivity $\sigma_2 = 5~{\rm S/m}$ and electrical permittivity $\varepsilon_2 = 81~\varepsilon_0$, which are typical for sea water~\cite{Martinez01}. 
The bottom layer represents a seabed with conductivity $\sigma_3 = 0.05~{\rm S/m}$ and permittivity $\varepsilon_3 = 15~\varepsilon_0$ \cite{Martinez01}.
The three SC cables are placed 1~m below the sea-seabed interface, as shown in Fig.~\ref{fig:Validation1}.
The center-to-center distance between adjacent cables is  $D = 85~{\rm mm}$.
Since the SC cables are touching each other, significant proximity effects between them are expected.

\subsubsection{Impedance Validation}
\begin{figure}[t]
\includegraphics[scale=0.9]{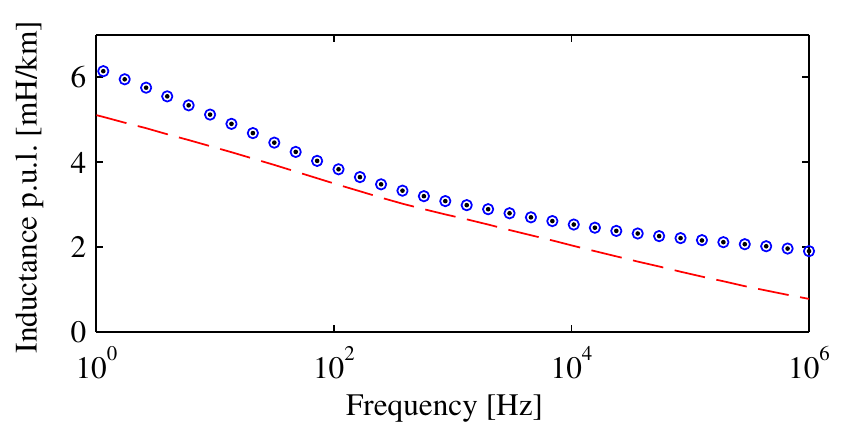}
\includegraphics[scale=0.9]{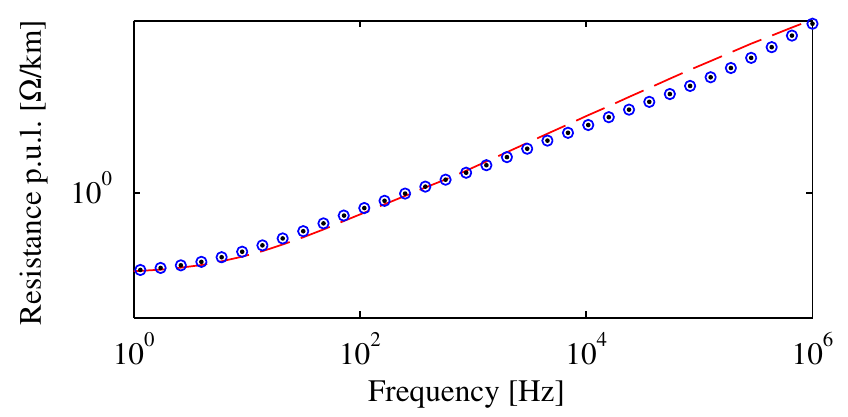}
\caption{Cable system of Sec.~\ref{sec:example1}: zero-sequence inductance (top panel), and resistance (bottom panel) obtained with MoM-SO ({\color{blue} ${\bf \mathlarger \circ}$}), FEM ($\cdot$), and analytic formulas ({\color{red} \xdash[.4em]~\xdash[.4em] }). Cable screens are left open.}
\label{fig:3sccable_open_zero_85mm}
\end{figure}
\begin{figure}[t h]
\includegraphics[scale=0.9]{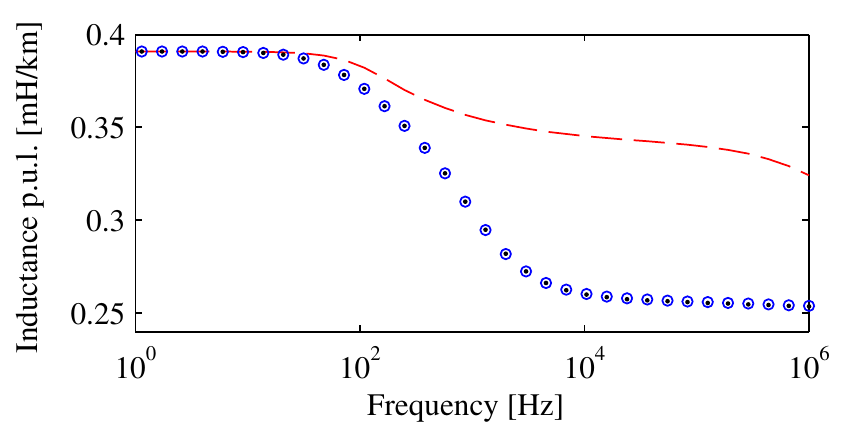}
\includegraphics[scale=0.9]{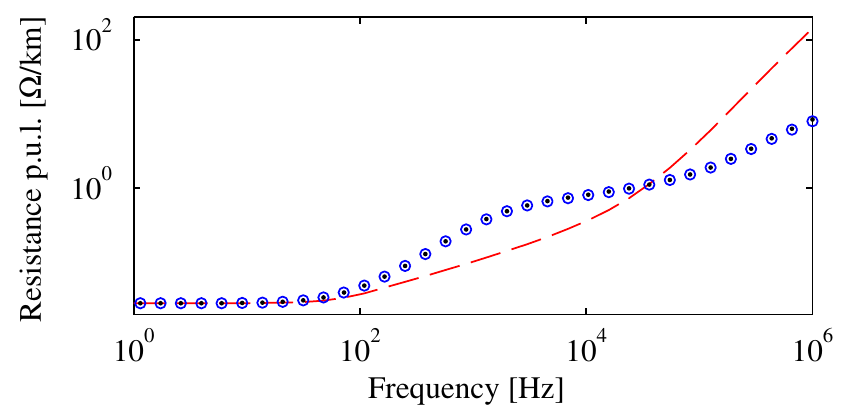}
\caption{As in Fig.~\ref{fig:3sccable_open_zero_85mm}, but when a positive-sequence excitation is applied to the core conductors.}
\label{fig:3sccable_open_positive_85mm}
\end{figure}

We first calculated the $6\times 6$ impedance matrix of the cable using the proposed MoM-SO approach.
In MoM-SO, the discretization parameters $N_p$ and $\widehat{N}$ were set to 4.
Next, we repeated the computation with a commercial FEM solver (COMSOL Multiphysics \cite{COMSOL}), following the approach presented in \cite{yin89}. 
In order to reproduce skin effect adequately, boundary layer elements were used to finely mesh the conductor edges.
In FEM, we also used the infinite element domain to truncate the surrounding medium. 
A total of 246,818 elements were required in the FEM simulation to mesh the cable and the surrounding medium.
Finally, we calculated the impedance of the cable with the  analytic formulas (cable constants \cite{Ame80}) that are implemented in most EMT tools.

In order to facilitate the comparison of the results obtained with MoM-SO, FEM, and analytic formulas, we reduced the $6 \times 6$ impedance matrix to a $ 3\times 3$ matrix by assuming that the screens of the cables are open at both ends\footnote{This assumption is made only to simplify the comparison of the cable parameters, and will not be used in the subsequent transient simulations.}, i.e. we assumed that  there is zero net current inside the screens.
Figures~\ref{fig:3sccable_open_zero_85mm} and~\ref{fig:3sccable_open_positive_85mm} show the cable inductance and resistance obtained with MoM-SO, FEM, and analytic formulas when zero- and positive-sequence excitations are applied to the core conductors. An excellent agreement between the results obtained with the proposed MoM-SO method and FEM can be observed. Analytic formulas instead return inaccurate parameters in both cases. In the case of positive sequence excitation (Fig.~\ref{fig:3sccable_open_positive_85mm}), inaccuracy is mainly attributed to the neglection of proximity effects. In the zero sequence case (Fig.~\ref{fig:3sccable_open_zero_85mm}) inaccuracy is due to proximity effects and the lack of an accurate multilayer ground model.

\subsubsection{Timing Comparison} 
The simulation times to calculate the impedance of the cable are summarized in Table~\ref{tab:Timing2}. 
FEM took 183~s per frequency to calculate the impedance of this cable.
On the otherhand, MoM-SO took just 0.10~s per frequency to calculate the impedance parameters with the same accuracy.

\begin{table}[t]
\centering
\caption{Example of Sec.~\ref{sec:example1}: CPU time required to compute the impedance at one frequency*}
\begin{tabular}{|c|c|c|c|}
\hline
{\bf Test Case} & {\bf MoM-SO}  & {\bf FEM}  & {\bf Speed-up}\\  \hline
Three-layers &  0.105 & 183 & 1743 X \\ \hline
\end{tabular} \\
\vskip 6pt
{*Simulations were run on a system with 16~GB memory \\
 and 3.40 GHz processor.}
\label{tab:Timing2} 
\end{table}

\subsubsection{Transient Simulation - Crossbonded Cable}
\label{sec:transient}

\begin{figure}[t]
\centering
\includegraphics[scale = 0.7]{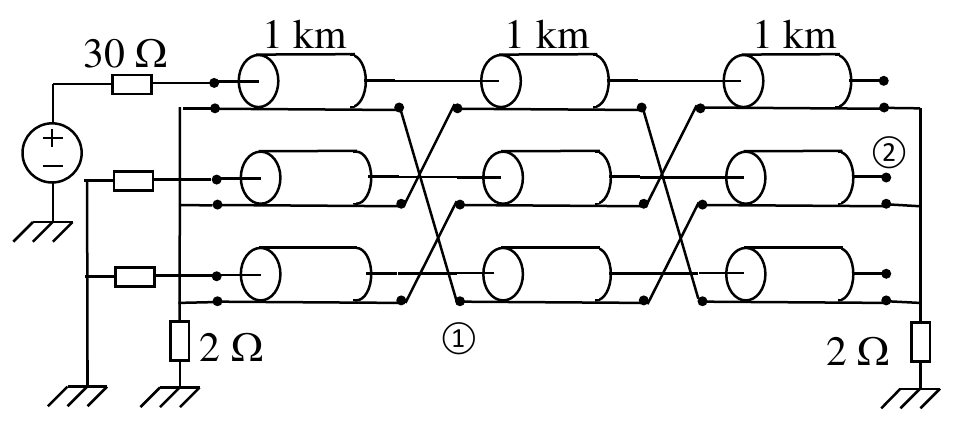}
\caption{Cross-bonded cable system setup.}
\label{fig:3sccable_transient_setup}
\end{figure}

\begin{figure}[t]
\centering
\includegraphics[scale=0.9]{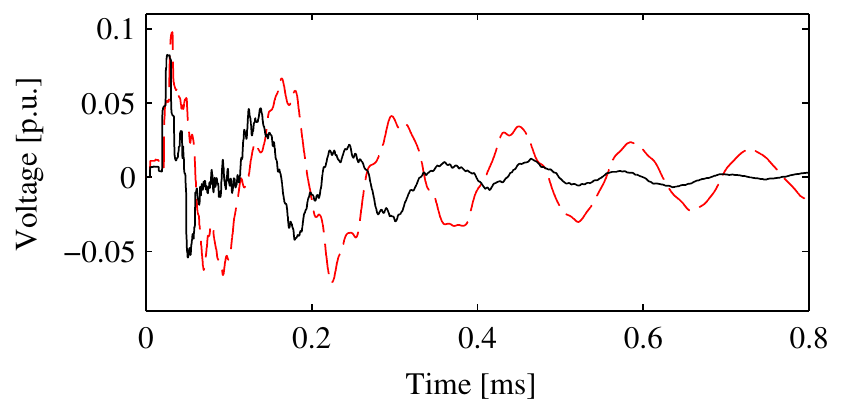}
\includegraphics[scale=0.9]{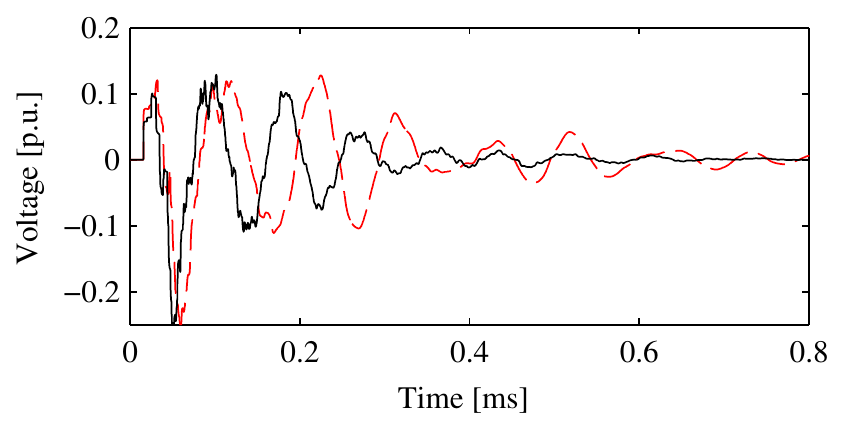}
\caption{Cable System of Sec.~\ref{sec:example1}: node 1 and node 2 voltages obtained with MoM-SO ({\color{black} \xdash[.8em]}) and  analytic formulas ({\color{red} \xdash[.4em]~\xdash[.4em] }) for the setup shown in Fig.~\ref{fig:3sccable_transient_setup}. }
\label{fig:3sccable_transient_results_prox}
\end{figure}

Next, we compare the transient waveform predicted with analytic formulas and MoM-SO.
We consider the setup in Fig.~\ref{fig:3sccable_transient_setup} where a unit step excitation is applied to the core conductor of the left-most SC cable in Fig.~\ref{fig:Validation1}.
We created two universal line models~\cite{Mor99} for the cable. The first model was derived from the impedance obtained with MoM-SO, while the second model was derived from the impedance calculated with analytic formulas. For both models, shunt admittance was calculated using the formulas from~\cite{Vel10}.
Figure~\ref{fig:3sccable_transient_results_prox} shows the transient voltages at node 1 and node 2 predicted with MoM-SO and analytic formulas.
The voltage waveforms predicted with analytic formulas significantly deviate from the waveforms predicted with MoM-SO, as a result of the neglection of proximity and multilayer ground effects. These results show how the proposed method can lead to more accurate transient results with respect to existing EMT tools, which are mostly based on analytic formulas.

\subsection{Example \# 2: Three Widely Separated Single-Core Cables }
\label{sec:example2}

\subsubsection{Geometry}
The first example showed the influence of proximity and multilayer ground effects on cable impedance and transient results.
In this example, we increase the separation between SC cables to 2~m, thereby minimizing proximity effects.
We compare three different ground models:
\begin{itemize}
\item three-layer air-sea-seabed model, where the surrounding medium is modelled as in Fig.~\ref{fig:Validation1};
\item two-layer air-sea model, where the presence of the seabed is neglected, and the sea layer extends to $y = -\infty$;
\item two-layer sea-seabed model, where the presence of air is neglected, and the sea layer extends to $y = \infty$.
\end{itemize}
The two-layer models represent what is currently possible with most EMT tools, which model ground as a single medium or as a two-layer medium~\cite{Saad96}.

\subsubsection{Grounded Screens}
\label{sec:impedancevalidation}

\begin{figure}[t]
\includegraphics[scale=0.9]{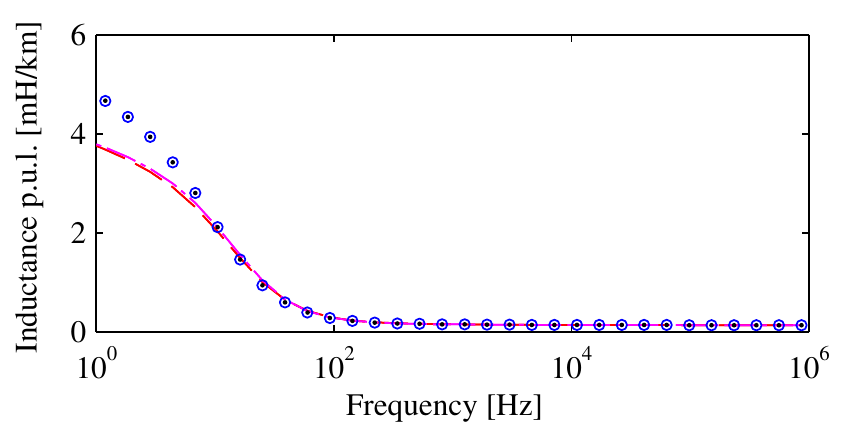}
\includegraphics[scale=0.9]{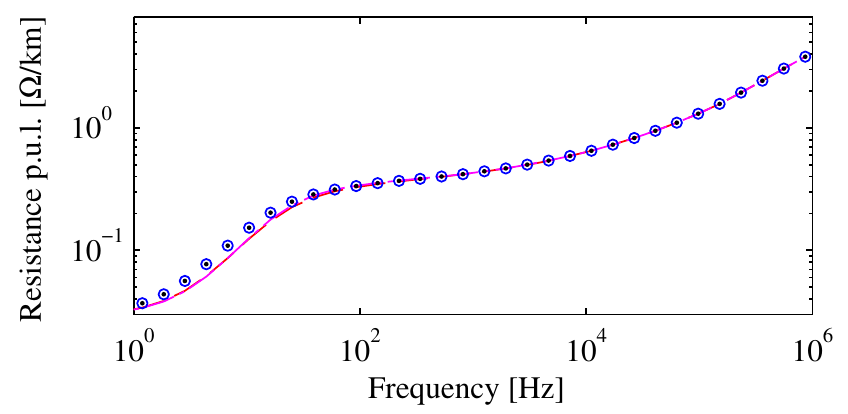}
\caption{Cable system of Sec.~\ref{sec:example2}: zero-sequence p.u.l. inductance (top panel) and resistance (bottom panel) obtained with the three-layer air-sea-seabed model in MoM-SO ({\color{blue} ${\bf \mathlarger \circ}$}), three-layer air-sea-seabed model in FEM ($\cdot$), two-layer air-sea model ({\color{red} \xdash[.4em]~\xdash[.4em] }), and two-layer sea-seabed model ({\color{magenta} \xdash[.4em]~\xdash[.1em]~\xdash[.4em]}). The screens are continuously grounded.
}
\label{fig:3sccable_grounded_zero}
\end{figure}
\begin{figure}[h]
\includegraphics[scale=0.9]{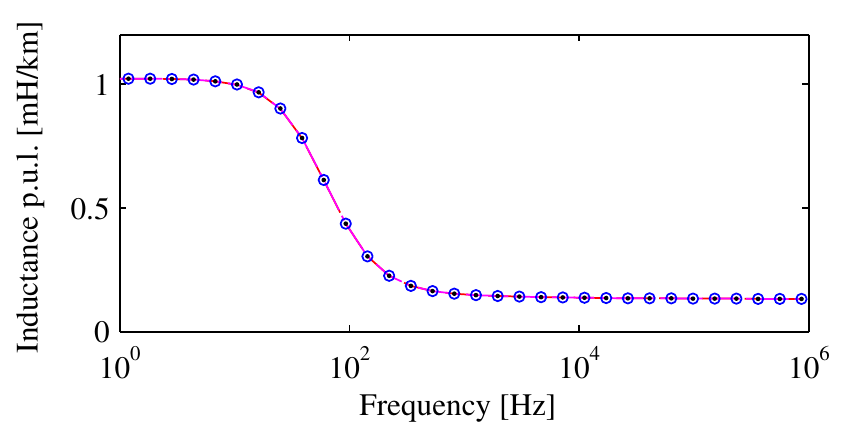}
\includegraphics[scale=0.9]{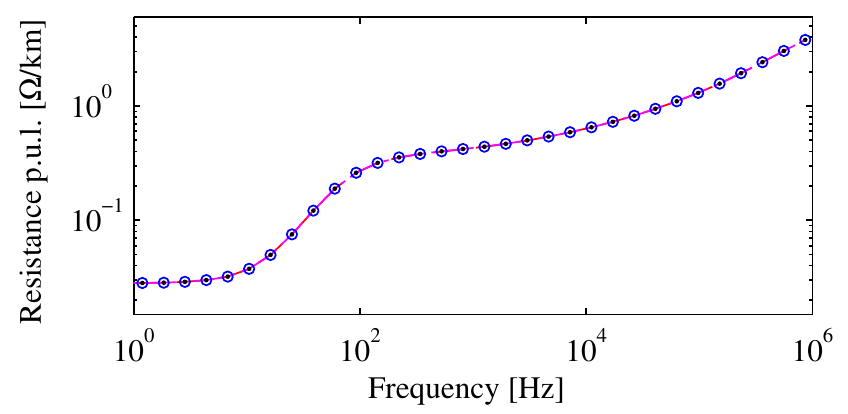}
\caption{As in Fig.~\ref{fig:3sccable_grounded_zero}, but when a positive-sequence excitation is applied to the core conductors.}
\label{fig:3sccable_grounded_positive}
\end{figure}

\begin{figure}[h t]
\includegraphics[scale=0.9]{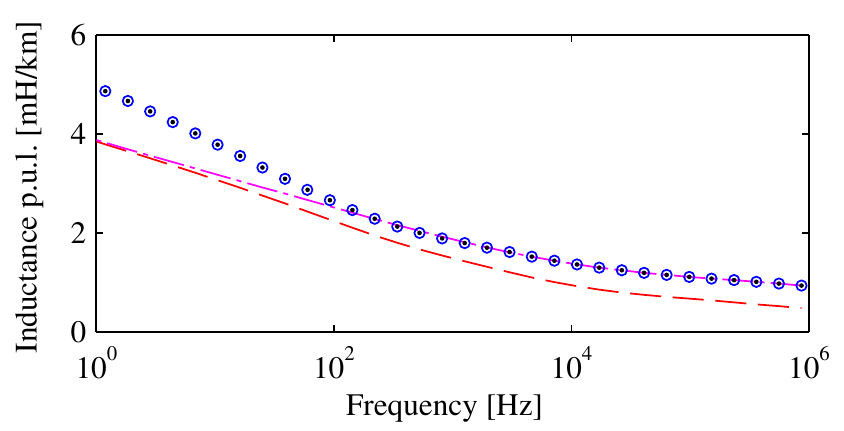}
\includegraphics[scale=0.9]{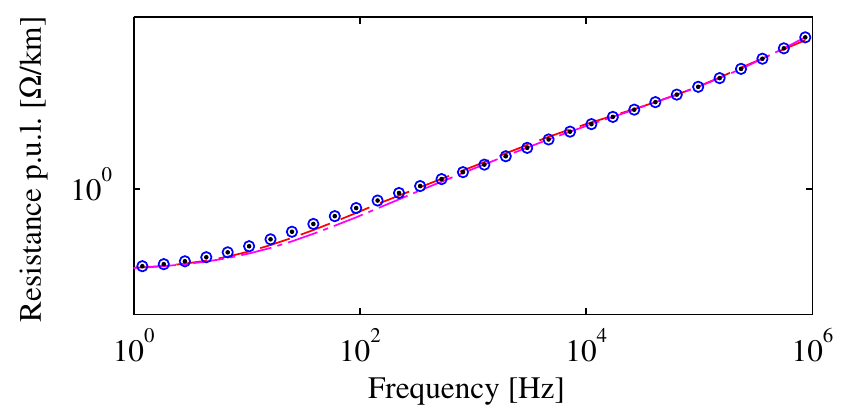}
\caption{Cable system of Sec.~\ref{sec:example2}: zero-sequence inductance (top panel), and resistance (bottom panel) obtained with the three-layer air-sea-seabed model in MoM-SO ({\color{blue} ${\bf \mathlarger \circ}$}), three-layer air-sea-seabed model in FEM ($\cdot$), two-layer air-sea model ({\color{red} \xdash[.4em]~\xdash[.4em] }), and two-layer sea-seabed model ({\color{magenta} \xdash[.4em]~\xdash[.1em]~\xdash[.4em]}). Cable screens are left open. }
\label{fig:3sccable_open_zero}
\end{figure}
\begin{figure}[h t]
\includegraphics[scale=0.9]{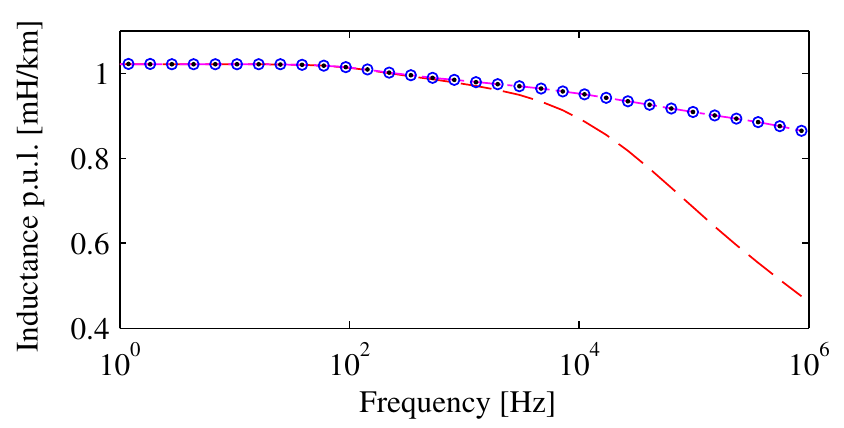}
\includegraphics[scale=0.9]{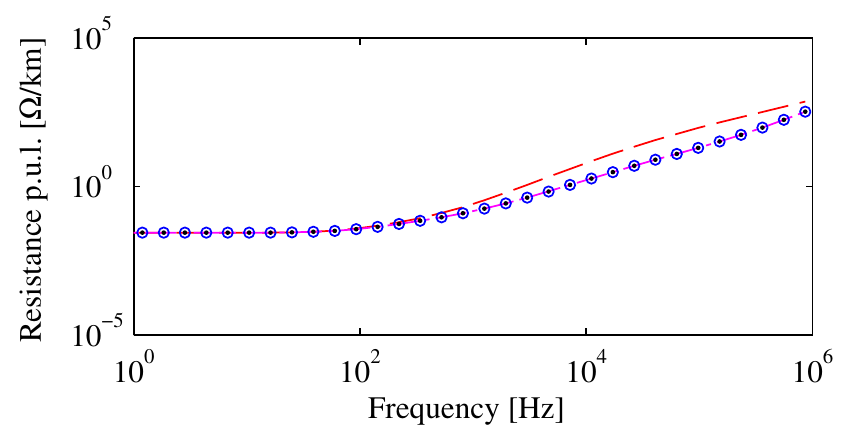}
\caption{As in Fig.~\ref{fig:3sccable_open_zero}, but when a  positive-sequence excitation is applied to the core conductors.}
\label{fig:3sccable_open_positive}
\end{figure}

We calculated the impedance of the cable using MoM-SO with the three ground models.
For validation, the computation was also performed with FEM in the three-layer case.
The $6 \times 6$ matrix was then reduced to a $ 3\times 3$ impedance matrix by assuming that the screens of the cable were continuously grounded.
Figures~\ref{fig:3sccable_grounded_zero} and~\ref{fig:3sccable_grounded_positive} show the p.u.l. inductance and p.u.l. resistance values obtained using both MoM-SO and FEM (three-layer model only) when zero- and positive-sequence excitations are applied to the core conductors.
Figures~\ref{fig:3sccable_grounded_zero} and~\ref{fig:3sccable_grounded_positive} show an excellent agreement between the impedance calculated with FEM and MoM-SO using the three-layer air-sea-seabed model, which validates the proposed approach.
On the other hand, we note that the two-layer models underestimate the zero-sequence inductance at low frequencies. 
These results confirm the superior accuracy of the proposed method with respect to existing techniques~\cite{Saad96,TPWRD3}.

\subsubsection{Open Screens}
We now assume that the cable screens are open.
Figures~\ref{fig:3sccable_open_zero} and~\ref{fig:3sccable_open_positive} show the zero- and positive-sequence inductance and resistance obtained by exciting core conductors with zero- and positive-sequence currents. The parameters obtained with MoM-SO match very well the reference FEM results.
The two-layer air-sea model produces inadequate results because at low frequencies the zero-sequence inductance is underestimated, and at high frequencies the positive-sequence inductance is underestimated.
The two-layer sea-seabed model is more accurate than the air-sea model, but still underestimates the zero-sequence inductance at low frequency. It should be noted that the two layer sea-seabed model cannot be utilized in some EMT tools  that require the conductivity of the top layer to be zero.

\subsubsection{Timing Comparison}
In MoM-SO, discretization parameters $N$ and $\widehat{N}$ were set to 4. In FEM, the cable and surrounding medium were meshed with 255,380 elements.
The timing results for both FEM and MoM-SO simulations are summarized in Table~\ref{tab:Timing1}. MoM-SO takes only 0.1~s per frequency, and is 1661 times faster than FEM, which requires almost three minutes per frequency.

\begin{table}[h t]
\centering
\caption{Example of Sec.~\ref{sec:example2}: CPU time required to compute the impedance at one frequency*}
\begin{tabular}{|c|c|c|c|}
\hline
{\bf Test Case} & {\bf MoM-SO}  & {\bf FEM}  & {\bf Speed-up}\\  \hline
Three-layers &  0.108 & 179 & 1661 X \\ \hline
\end{tabular} \\
\vskip 6pt
{*Simulations were run on a system with 16~GB memory \\
 and 3.40 GHz processor.}
\label{tab:Timing1} 
\end{table}

\subsubsection{Transient Simulation - Crossbonded cable}
\label{sec:transient1}

\begin{figure}[ht]
\centering
\includegraphics[scale=0.9]{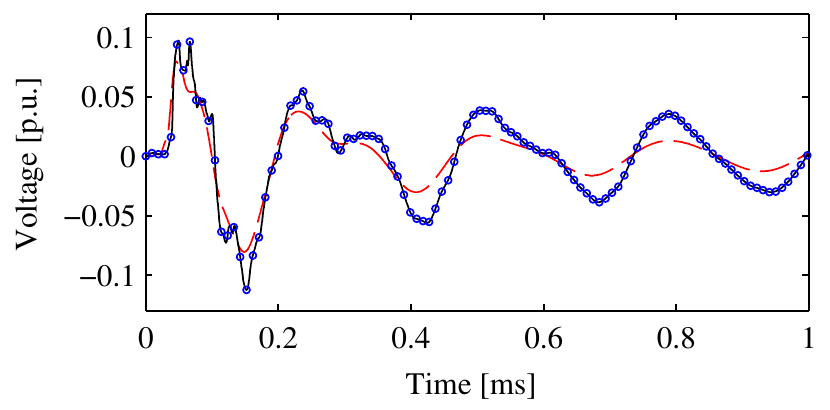}
\includegraphics[scale=0.9]{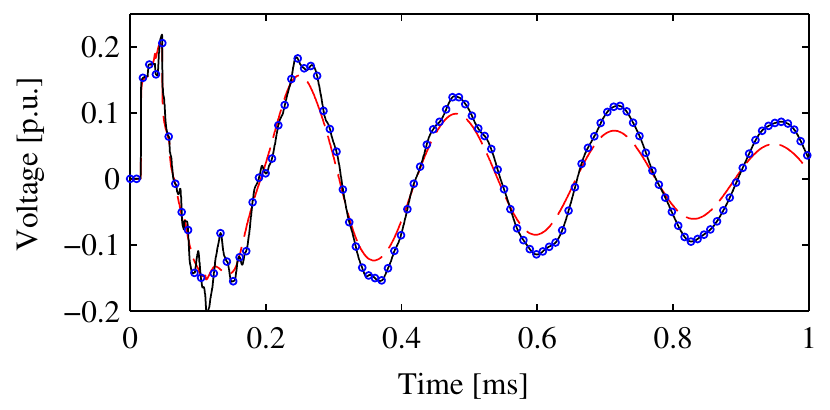}
\caption{Example of Sec.\ref{sec:example2}: voltages predicted at nodes 1 (top panel) and 2 (bottom panel) of the configuration in Fig.~\ref{fig:3sccable_transient_setup}. Plots compare the results obtained with three different ground models: three-layer air-sea-seabed({\color{black} \xdash[.8em]}), two-layer air-sea ({\color{red} \xdash[.4em]~\xdash[.4em] }), and two-layer sea-seabed model ({\color{blue} ${\bf \mathlarger \circ}$}). }
\label{fig:3sccable_transient_results}
\end{figure}

Finally, we compute the transient voltages excited by a unit step voltage applied to the 
three SC cables in cross-bonded configuration (see Fig.~\ref{fig:3sccable_transient_setup}). 
Figure~\ref{fig:3sccable_transient_results} shows the transient voltages at nodes 1 and 2.
The results show that there is up to 40\% deviation between the results obtained with the air-sea model and the air-sea-seabed model. 
In this case, the two-layer sea-seabed model returns accurate results. However, if the water depth is reduced, this model becomes inaccurate. Since no two-layer model is accurate under all possible cases, their use requires the EMT engineer to understand which one is more appropriate for a certain cable. With MoM-SO, this dilemma is avoided, and accurate cable parameters are computed in less than a second, making cable modeling a straightforward and less error-prone task.
\section{Conclusion}

A multilayer ground model was proposed for the MoM-SO approach for cable impedance calculation.
With the proposed model, the non-uniformity of the medium which surrounds submarine and underground cables can be accurately taken into account. 
Numerical results show that the proposed method leads to better transient predictions than analytic formulas currently used in most electromagnetic transient simulators. While the level of achievable accuracy is comparable to a finite elements analysis, MoM-SO is more than 1000 times faster than finite elements. Moreover, it is simpler to use, since it is fully automated and avoids meshing-related issues. We believe that these improvements make accurate cable modeling a simpler task for both transients experts as well as power engineers in general.

\section{Acknowledgement}
Authors thank Dr. Bj\o rn Gustavsen (SINTEF Energy Research, Norway) for providing the test cases in Sec.~\ref{sec:Examples}.

\bibliography{IEEEabrv,biblio}

\end{document}